\documentclass [onecolumn]{mn2e}

\title[Eccentricity evolution in HTS]{Eccentricity evolution in
hierarchical triple systems with eccentric outer binaries}
\author[Nikolaos Georgakarakos]{Nikolaos Georgakarakos\\School of Mathematics, Edinburgh
University\\Mayfield Road, Edinburgh EH9 3JZ\\
email: ng@maths.ed.ac.uk}
\date{}

\usepackage{graphicx}

\begin{document}
\maketitle

\begin{abstract}

We develop a technique for estimating the inner
eccentricity in hierarchical triple systems, with the inner
orbit being initially circular ,while the outer one is eccentric.  We
consider coplanar systems with well separated components and comparable
masses.  The derivation of short period terms is based on an expansion of the rate of
change of the Runge-Lenz vector.  Then, the short period terms are
combined with secular terms, obtained by
means of canonical perturbation theory.   The validity of the 
theoretical equations is tested by numerical integrations of the
full equations of motion.
\end{abstract}


\section{INTRODUCTION}

A hierarchical triple system consists of a binary system and a third body on a
wider orbit.  The motion of such a system can be pictured as the motion
of two binaries: the binary itself (inner binary) and the binary which
consists of the third body and the centre of mass of the binary (outer binary).  Hierarchical triple systems are widely present in the
galactic field and in star clusters and studying the dynamical evolution of such
systems is a key to understanding a number of issues in astronomy and
astrophysics.  Sometimes, for example, the inner pairs in triple
stellar systems are {\sl close}
binary systems, i.e. the separation between the components is
comparable to the radii of the bodies.  In these circumstances, the
behaviour of the inner binary can depend very sensitively on the
separation of its components and this in turn is affected by the third
body.  Thus, a slight change in the separation of the binary
stars can cause drastic changes in processes such as tidal friction
and dissipation, mass transfer and  mass loss due to a stellar wind,
which may result in changes in stellar
structure and evolution (e.g. Kiseleva, Eggleton ${\&}$ Mikkola 1998).  Eventually, these physical changes
can affect the dynamics of the whole triple system.  But even in
systems with well-separated inner binary components, the perturbation
of the third body can have a devastating effect on the 
triple system as a whole (e.g. disruption of the system).     

For most hierarchical triple stars, the period ratio
${X}$ is of the order of 100 and these systems are probably very stable
dynamically.  However, there are systems with much smaller period
ratios, like the system HD 109648 with ${X=22}$ (Jha et al. 2000), the
${\lambda}$ Tau system, with
\begin{math}
X=8.3
\end{math}
(Fekel ${\&}$ Tomkin 1982)
and the CH Cyg system with
\begin{math}
X=7.0
\end{math}
(Hinkle et al. 1993). 

In a previous paper (Georgakarakos 2002) we derived a formula for the
inner eccentricity in hierarchical triple systems with coplanar and
initially circular orbits.  Now, the calculation is extended to
systems with eccentric outer binaries (the inner orbit is still
considered to be initially circular).

\section{THEORY}
\label{theo}
We derive expressions for the short period and
secular evolution of the inner eccentricity.  Both short period and
secular terms will be obtained as previously (Georgakarakos 2002),
i.e.  by using the definition of the Runge-Lenz vector for the former
and by means of canonical perturbation theory for the latter.  
Again, at any moment of the evolution of the system, the eccentricity
is considered to consist of a short period and a long period (secular)
component, i.e. ${e=e_{{\rm short}}+e_{{\rm sec}}}$ (one can picture this by
recalling the expansion of the disturbing function in solar system
dynamics, where the perturbing potential is given as a sum of an
infinite number of cosines of various frequencies).  Thus, the
eccentricity being initially zero implies that ${e_{{\rm
short}}=-e_{{\rm sec}}}$ (initially).

Finally, in this paper, we will be
concentrating on systems with well separated components and comparable
masses.  Therefore, while developing the theoretical model in the next
sections, we will consider ${X}$ to be large (or any equivalent form
of that assumption). 

\subsection{Calculation of the short-period contribution to the eccentricity}	
\label{s1}
\begin{figure}
\begin{center}
\includegraphics[width=70mm,height=50mm]{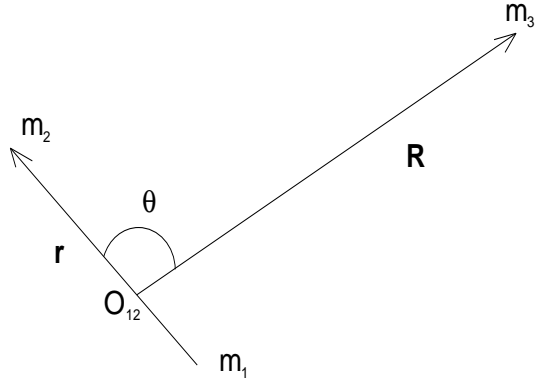}
\caption[]{The Jacobi formulation.  The point ${O_{12}}$ is the centre
of mass of the inner binary.}
\label{jacform}
\end{center}
\end{figure}

First, we calculate the short-period terms.  The motion of the system can be studied using the Jacobi decomposition
of the three-body problem (Fig. \ref{jacform}).  In that context, the equation of motion of
the inner binary is:
\begin{equation}
\ddot{\bmath {r}}=-G(m_{1}+m_{2})\frac{\bmath{r}}{r^{3}}+\bmath{F},
\label{eqmo}
\end{equation}
where ${\bmath{F}}$, the perturbation to the inner binary motion, is  
\begin{equation}
\bmath{F}=Gm_{3}(\frac{\bmath{R}-\mu_{1}\bmath{r}}{|\bmath{R}-\mu_{1}\bmath{r}|^{3}}-\frac{\bmath{R}+\mu_{2}\bmath{r}}{|\bmath{R}+\mu_{2}\bmath{r}|^{3}})=Gm_{3}\frac{\partial}{\partial{\bmath{r}}}(\frac{1}{\mu_{1}|\bmath{R}-\mu_{1}\bmath{r}|}+\frac{1}{\mu_{2}|\bmath{R}+\mu_{2}\bmath{r}|})
\label{potenc1}
\end{equation}
with
\begin{displaymath}
\mu_{{\rm i}}=\frac{m_{{\rm i}}}{m_{1}+m_{2}}, \hspace{0.5 cm} i=1,2.
\end{displaymath}
Now, since the third star is at considerable distance from
the inner binary, implying that 
\begin{math}
r/R 
\end{math}
is small, the inverse distances in equation (\ref{potenc1}) can be expressed as:
\begin{displaymath}
\frac{1}{|\bmath{R}-\mu_{1}\bmath{r}|}=\frac{1}{R}\sum^{\infty}_{n=o}
\left( \frac{\mu_{1}r}{R} \right) ^{n}P_{{\rm n}}(\cos{\theta})
\end{displaymath}
and
\begin{displaymath}
\frac{1}{|\bmath{R}+\mu_{2}\bmath{r}|}=\frac{1}{R}\sum^{\infty}_{n=o}
\left(- \frac{\mu_{2}r}{R} \right)^{n}P_{{\rm n}}(\cos{\theta}),
\end{displaymath}
where ${P_{{\rm n}}}$ are the Legendre polynomials and ${\theta}$ is the
angle between the vectors ${\bmath{r}}$ and ${\bmath{R}}$.  Expanding to third
order, the perturbation becomes

\begin{equation}
\bmath{F}=Gm_{3}\frac{\partial}{\partial{\bmath{r}}}\left(\frac{3}{2}\frac{(\bmath{r}\cdot
\bmath{R})^{2}}{R^{5}}-\frac{1}{2}\frac{r^{2}}{R^{3}}-\frac{5(\mu_{2}^{2}-\mu_{1}^{2})}{2}\frac{(\bmath{r}\cdot\bmath{R})^{3}}{R^{7}}+\frac{3(\mu_{2}^{2}-\mu_{1}^{2})}{2}\frac{r^{2}(\bmath{r}\cdot\bmath{R})}{R^{5}}\right).
\label{fpert}
\end{equation}
The first two terms in the above equation come from the quadrupole
term (${P_{2}}$), while the other two come from the octupole term
(${P_{3}}$).  

Using now the definition of the eccentric vector,
i.e. the vector which has the same direction as the radius vector to
the pericentre and whose magnitude is equal to the eccentricity of the
orbit, we can obtain an expression for the inner eccentricity.  The
inner eccentric vector
${\bmath{e}_{1}}$ is given by
\begin{equation}
\bmath{e}_{1}=-\frac{\bmath{r}}{r}+\frac{1}{\mu}(\dot{\bmath{r}}\bmath{\times}\bmath{h}),
\label{ecve}
\end{equation}
where
\begin{math}
\bmath{h}=\bmath{r}\bmath{\times}\dot{\bmath{r}}
\end{math}
and
\begin{math}
\mu=G(m_{1}+m_{2}).
\end{math} 
Differentiating equation (\ref{ecve}) and substituting for
${\bmath{F}}$ (we neglect the term
${\bmath{r}\cdot\dot{\bmath{r}}}$ because, for the
applications discussed in this paper, is expected to be  small and of ${O(e)}$), we obtain:
\begin{eqnarray}
\dot{\bmath{e}}_{1} & = & \frac{Gm_{3}}{\mu R^{3}}\left[\left(6\frac{(\bmath{r}\cdot\bmath{R})(\dot{\bmath{r}}\cdot\bmath{R})}{R^{2}}-15(\mu_{2}^{2}-\mu_{1}^{2})\frac{(\bmath{r}\cdot\bmath{R})^{2}(\dot{\bmath{r}}\cdot\bmath{R})}{R^{4}}+3(\mu_{2}^{2}-\mu_{1}^{2})\frac{r^{2}(\dot{\bmath{r}}\cdot\bmath{R})}{R^{2}}\right)\bmath{r}+\right.\nonumber\\
& & \left.+\left(r^{2}-3\frac{(\bmath{r}\cdot\bmath{R})^{2}}{R^{2}}+\frac{15}{2}(\mu_{2}^{2}-\mu_{1}^{2})\frac{(\bmath{r}\cdot\bmath{R})^{3}}{R^{4}}-\frac{9}{2}(\mu_{2}^{2}-\mu_{1}^{2})\frac{r^{2}(\bmath{r}\cdot\bmath{R})}{R^{2}}\right)\dot{\bmath{r}}\right].\label{roo5}
\end{eqnarray}
Now, the Jacobi vectors can be represented approximately in polar form as
\begin{math}
\bmath{r}=a_{1}(\cos{n_{1}t},\sin{n_{1}t})
\end{math}
and
\begin{math}
\bmath{R}=R(\cos{(f+\varpi)},\\ \sin{(f+\varpi)})
\end{math}
, where ${a_{1}}$ is the
semi-major axis of the inner orbit, ${n_{1}}$ is the mean motion of
the inner orbit,
\begin{math}
R=\frac{a_{2}(1-e^{2})}{1+e\cos{f}}
\end{math}
, ${a_{2}}$ the outer semi-major axis, ${e}$ the outer eccentricity,
and ${f}$ and ${\varpi}$ are the true
anomaly and longitude of pericentre of the outer orbit.
After integrating, the components ${x_{1}}$ and ${y_{1}}$ of the
eccentric vector become (expanding in powers of
\begin{math}
\frac{1}{X}
\end{math}
and retaining the four leading terms):

\begin{eqnarray}
x_{1}(t) & = &
\frac{m_{3}}{M}\frac{1}{X^{2}(1-e^{2})^{3}}\left( P_{{\rm x21}}(t)+\frac{1}{X}P_{{\rm x22}}(t)+m_{*}X^{\frac{1}{3}}P_{{\rm x31}}(t)+m_{*}\frac{1}{X^{\frac{2}{3}}}P_{{\rm x32}}(t)\right)+C_{{\rm
x}_{1}}\label{e11}\\
y_{1}(t) & = &
\frac{m_{3}}{M}\frac{1}{X^{2}(1-e^{2})^{3}}\left(P_{{\rm y21}}(t)+\frac{1}{X}P_{{\rm y22}}(t)+m_{*}X^{\frac{1}{3}}P_{{\rm y31}}(t)+m_{*}\frac{1}{X^{\frac{2}{3}}}P_{{\rm y32}}(t)\right)+C_{{\rm
y}_{1}}\label{e12}
\end{eqnarray}
where
\begin{eqnarray}
P_{{\rm x21}}(t) & = & (1+e\cos{f})^{3}[-\frac{1}{2}\cos{n_{1}t}+\frac{1}{4}\cos{(3n_{1}t-2f-2\varpi)}+\frac{9}{4}\cos{(n_{1}t-2f-2\varpi)}]\\
P_{{\rm x22}}(t) & = &
\frac{(1+e\cos{f})^{4}}{(1-e^{2})^{\frac{3}{2}}}
\left[\frac{9}{2}\cos{(n_{1}t-2f-2\varpi)}+\frac{1}{6}\cos{(3n_{1}t-2f-2\varpi)}+e[-\frac{3}{4}\cos{(n_{1}t-f)}+\frac{3}{4}\cos{(n_{1}t+f)}+\right.\nonumber\\
& & \left.+\frac{45}{8}\cos{(n_{1}t-3f-2\varpi)}+\frac{5}{24}\cos{(3n_{1}t-3f-2\varpi)}-\frac{9}{8}\cos{(n_{1}t-f-2\varpi)}-\frac{1}{24}\cos{(3n_{1}t-f-2\varpi)}]\right]\\
P_{{\rm x31}}(t) & = &
(1-e^{2})^{\frac{1}{2}}\left[\frac{15}{16}\cos{(f+\varpi)}+\frac{15}{32}e\cos{(2f+\varpi)}+e^{2}[\frac{45}{32}\cos{(f+\varpi)}-\frac{75}{64}\cos{(f-\varpi)}+\frac{5}{64}\cos{(3f+\varpi)}]+\right.\nonumber\\
& & +\left.e^{3}[\frac{45}{128}\cos{(2f-\varpi)}-\frac{45}{128}\cos{(2f+\varpi)}]+e^{4}[\frac{5}{32}\cos{(3f+\varpi)}-\frac{5}{32}\cos{(3f-\varpi)}]\right]\\
P_{{\rm x32}}(t) & = & \frac{(1+e\cos{f})^{4}}{(1-e^{2})}
[\frac{3}{32}\cos{(2n_{1}t-f-\varpi)}-\frac{45}{32}\cos{(2n_{1}t-3f-3\varpi)}-\frac{15}{64}\cos{(4n_{1}t-3f-3\varpi)}]\\
P_{{\rm y21}}(t) & = & (1+e\cos{f})^{3}[-\frac{1}{2}\sin{n_{1}t}+\frac{1}{4}\sin{(3n_{1}t-2f-2\varpi)}-\frac{9}{4}\sin{(n_{1}t-2f-2\varpi)}]\\
P_{{\rm y22}}(t) & = & \frac{(1+e\cos{f})^{4}}{(1-e^{2})^{\frac{3}{2}}}
\left[-\frac{9}{2}\sin{(n_{1}t-2f-2\varpi)}+\frac{1}{6}\sin{(3n_{1}t-2f-2\varpi)}+e[-\frac{3}{4}\sin{(n_{1}t-f)}+\frac{3}{4}\sin{(n_{1}t+f)}-\right.\nonumber\\
& & \left.-\frac{45}{8}\sin{(n_{1}t-3f-2\varpi)}+\frac{5}{24}\sin{(3n_{1}t-3f-2\varpi)}+\frac{9}{8}\sin{(n_{1}t-f-2\varpi)}-\frac{1}{24}\sin{(3n_{1}t-f-2\varpi)}]\right]\\
P_{{\rm y31}}(t) & = &
(1-e^{2})^{\frac{1}{2}}\left[\frac{15}{16}\sin{(f+\varpi)}+\frac{15}{32}e\sin{(2f+\varpi)}+e^{2}[\frac{45}{32}\sin{(f+\varpi)}+\frac{75}{64}\sin{(f-\varpi)}+\frac{5}{64}\sin{(3f+\varpi)}]+\right.\nonumber\\
& & \left.+e^{3}[-\frac{45}{128}\sin{(2f-\varpi)}-\frac{45}{128}\sin{(2f+\varpi)}]+e^{4}[\frac{5}{32}\sin{(3f+\varpi)}+\frac{5}{32}\sin{(3f-\varpi)}]\right]\\
P_{{\rm y32}}(t) & = & \frac{(1+e\cos{f})^{4}}{(1-e^{2})}[\frac{3}{32}\sin{(2n_{1}t-f-\varpi)}+\frac{45}{32}\sin{(2n_{1}t-3f-3\varpi)}-\frac{15}{64}\sin{(4n_{1}t-3f-3\varpi)}]
\end{eqnarray}
and
\begin{displaymath}
m_{*}=\frac{m_{2}-m_{1}}{(m_{1}+m_{2})^{\frac{2}{3}}M^{\frac{1}{3}}}.
\end{displaymath}
${M}$ is the total mass of the system and ${C_{{\rm x}_{1}}}$ and
${C_{{\rm y}_{1}}}$ are
constants of integration.  The semi-major axes, mean motions, outer
eccentricity and longitude of pericentre were
treated as constants in the above calculation.
It should be mentioned here that in the expressions for ${P_{{\rm
x31}}(t)}$ and ${P_{{\rm y31}}(t)}$ terms proportional to ${f}$
appeared.  To eliminate ${f}$ from our expressions, we used the
following series expansion (Murray ${\&}$ Dermott 1999)
\begin{equation}
f=l+2e\sin{f}-\frac{3}{4}e^{2}\sin{2f}+\frac{1}{3}e^{3}\sin{3f}+O(e^{4}),\label{expa}
\end{equation}
where ${l}$ is the mean anomaly, i.e.  we replaced ${f}$ with the
periodic part of the above equation.

\subsection{Calculation of the secular contribution to the eccentricity}	
\label{s2}
In order to derive the long-term modulation of the system, we use a
Hamiltonian which is averaged over the inner and outer orbital periods
by means of the Von Zeipel method.  Secular terms cannot be obtained
by the method of section \ref{s1}, because, as we just saw, those terms appear as a linear function of time in the
expansion of the eccentric vector and therefore, they are valid for limited time.

The doubly averaged Hamiltonian for coplanar orbits is (Marchal 1990):

\begin{eqnarray}
H & = &-\frac{Gm_{1}m_{2}}{2a_{{\rm
S}}}-\frac{G(m_{1}+m_{2})m_{3}}{2a_{{\rm T}}}+Q_{1}+Q_{2}+Q_{3}, \label{hamilto} \\
\mbox{where}\nonumber\\
Q_{1} & = &
-\frac{1}{8}\frac{Gm_{1}m_{2}m_{3}a^{2}_{{\rm
S}}}{(m_{1}+m_{2})a^{3}_{{\rm T}}(1-e^{2}_{{\rm T}})^{\frac{3}{2}}}(2+3e^{2}_{{\rm S}}), \\
Q_{2} & = &
\frac{15Gm_{1}m_{2}m_{3}(m_{1}-m_{2})a^{3}_{{\rm S}}e_{{\rm S}}e_{{\rm
T}}}{64(m_{1}+m_{2})^{2}a^{4}_{{\rm T}}(1-e^{2}_{{\rm T}})^{\frac{5}{2}}}\cos{(g_{{\rm S}}-g_{{\rm T}})}(4+3e^{2}_{{\rm S}}),\label{marq}\\
Q_{3} & = &
-\frac{15}{64}\frac{Gm_{1}m_{2}m_{3}^{2}a_{{\rm
S}}^{\frac{7}{2}}e_{{\rm S}}^{2}(1-e_{{\rm
S}}^{2})^{\frac{1}{2}}}{(m_{1}+m_{2})^{\frac{3}{2}}M^{\frac{1}{2}}a_{{\rm
T}}^{\frac{9}{2}}(1-e_{{\rm T}}^{2})^{3}}[5(3+2e_{{\rm
T}}^{2})+3e^{2}_{{\rm T}}\cos{2(g_{{\rm S}}-g_{{\rm T}})}].
\end{eqnarray}
The subscripts S and T refer to the inner and outer long period
orbits respectively, while ${g}$ is used to denote longitude of pericentre.  The first term in the Hamiltonian is the Keplerian energy of the inner
binary, the second term is the Keplerian energy of the outer binary,
while the other three terms represent the interaction between the two
binaries.  The ${Q_{1}}$ term comes from the ${P_{2}}$ Legendre
polynomial, the ${Q_{2}}$ term comes from the ${P_{3}}$
Legendre polynomial and the ${Q_{3}}$ term arises from the canonical transformation.

By using Hamilton's equations, we can now derive the averaged
equations of motion of the system.  Hence,
\begin{eqnarray}
\frac{{\rm d}x_{{\rm S}}}{{\rm d}\tau} & = &
\frac{5}{16}\alpha\frac{e_{{\rm T}}}{(1-e^{2}_{{\rm
T}})^\frac{5}{2}}(1-e^{2}_{{\rm S}})^{\frac{1}{2}}[(4+3e^{2}_{{\rm S}})\sin{g_{{\rm T}}}+6(x_{{\rm S}}y_{{\rm S}}\cos{g_{{\rm T}}}+y_{{\rm
S}}^{2}\sin{g_{{\rm T}}})]-[\frac{(1-e^{2}_{{\rm S}})^{\frac{1}{2}}}{(1-e^{2}_{{\rm
T}})^{\frac{3}{2}}}+\frac{25}{8}\gamma\frac{3+2e^{2}_{{\rm
T}}}{(1-e^{2}_{{\rm T}})^{3}}(1-\nonumber\\
& &
-\frac{3}{2}e^{2}_{{\rm S}})]y_{{\rm
S}}+\frac{15}{8}\gamma\frac{e^{2}_{{\rm T}}}{(1-e^{2}_{{\rm
T}})^{3}}[y_{{\rm S}}\cos{2g_{{\rm T}}}-x_{{\rm S}}\sin{2g_{{\rm T}}}-\frac{y_{{\rm S}}}{2}(x^{2}_{{\rm
S}}+3y^{2}_{{\rm S}})\cos{2g_{{\rm T}}}+x_{{\rm S}}(x^{2}_{{\rm S}}+2y^{2}_{{\rm S}})\sin{2g_{{\rm T}}}]\\
\frac{{\rm d}y_{{\rm S}}}{{\rm d}\tau} & = &-\frac{5}{16}\alpha\frac{e_{{\rm T}}}{(1-e^{2}_{{\rm
T}})^\frac{5}{2}}(1-e^{2}_{{\rm S}})^{\frac{1}{2}}[(4+3e^{2}_{{\rm
S}})\cos{g_{{\rm T}}}+6(x_{{\rm S}}y_{{\rm S}}\sin{g_{{\rm T}}}+x_{{\rm
S}}^{2}\cos{g_{{\rm T}}})]+[\frac{(1-e^{2}_{{\rm S}})^{\frac{1}{2}}}{(1-e^{2}_{{\rm
T}})^{\frac{3}{2}}}+\frac{25}{8}\gamma\frac{3+2e^{2}_{{\rm
T}}}{(1-e^{2}_{{\rm T}})^{3}}(1-\nonumber\\
& &
-\frac{3}{2}e^{2}_{{\rm S}})]x_{{\rm
S}}+\frac{15}{8}\gamma\frac{e^{2}_{{\rm T}}}{(1-e^{2}_{{\rm
T}})^{3}}[x_{{\rm S}}\cos{2g_{{\rm T}}}+y_{{\rm S}}\sin{2g_{{\rm T}}}-\frac{x_{{\rm S}}}{2}(y^{2}_{{\rm
S}}+3x^{2}_{{\rm S}})\cos{2g_{{\rm T}}}-y_{{\rm S}}(y^{2}_{{\rm S}}+2x^{2}_{{\rm S}})\sin{2g_{{\rm T}}}]\\
\frac{{\rm d}g_{{\rm T}}}{{\rm d}\tau} & = & \frac{\beta
(2+3e^{2}_{{\rm S}})}{2(1-e^{2}_{{\rm T}})^{2}}-\frac{5}{16}\frac{\alpha \beta
(1+4e^{2}_{{\rm T}})}{e_{{\rm T}}(1-e^{2}_{{\rm
T}})^{3}}(4+3e^{2}_{{\rm S}})(x_{{\rm S}}\cos{g_{{\rm T}}}+y_{{\rm S}}\sin{g_{{\rm T}}})+\frac{5}{8}\beta\gamma\frac{(1-e^{2}_{{\rm S}})^{\frac{1}{2}}}{(1-e^{2}_{{\rm
T}})^{\frac{7}{2}}}[5e^{2}_{{\rm S}}(11+4e^{2}_{{\rm
T}})+3(1+2e^{2}_{{\rm T}})\times\nonumber\\
& & \times((x^{2}_{{\rm S}}-y^{2}_{{\rm S}})\cos{2g_{{\rm
T}}}+2x_{{\rm S}}y_{{\rm S}}\sin{2g_{{\rm T}}})]\label{gtd}\\
\frac{{\rm d}e_{{\rm T}}}{{\rm d}\tau} & = & \frac{5}{16}\frac{\alpha
\beta}{(1-e^{2}_{{\rm T}})^{2}}(4+3e^{2}_{{\rm S}})(y_{{\rm
S}}\cos{g_{{\rm T}}}-x_{{\rm S}}\sin{g_{{\rm T}}})-\frac{15}{8}\beta\gamma\frac{e_{{\rm
T}}(1-e^{2}_{{\rm S}})^{\frac{1}{2}}}{(1-e^{2}_{{\rm T}})^{\frac{5}{2}}}(2x_{{\rm S}}y_{{\rm S}}\cos{2g_{{\rm T}}}-(x^{2}_{{\rm
S}}-y^{2}_{{\rm S}})\sin{2g_{{\rm T}}})
\end{eqnarray}
where
\begin{displaymath}
x_{{\rm S}}=e_{{\rm S}}\cos{g_{{\rm S}}},\hspace{0.5cm}y_{{\rm S}}=e_{{\rm
S}}\sin{g_{{\rm S}}},
\end{displaymath}
\begin{displaymath}
\alpha =\frac{m_{1}-m_{2}}{m_{1}+m_{2}}\frac{a_{{\rm S}}}{a_{{\rm T}}},\hspace{0.2cm}\beta
=\frac{m_{1}m_{2}M^{\frac{1}{2}}}{m_{3}(m_{1}+m_{2})^{\frac{3}{2}}}(\frac{a_{{\rm
S}}}{a_{{\rm T}}})^{\frac{1}{2}},\hspace{0.2cm}\gamma=\frac{m_{3}}{M^{\frac{1}{2}}(m_{1}+m_{2})^{\frac{1}{2}}}(\frac{a_{{\rm
S}}}{a_{{\rm T}}})^{\frac{3}{2}}\hspace{0.5cm}
\mbox{and}\hspace{0.5 cm}
{\rm d}\tau=\frac{3}{4}\frac{G^{\frac{1}{2}}m_{3}a^{\frac{3}{2}}_{{\rm
S}}}{a^{3}_{{\rm T}}(m_{1}+m_{2})^{\frac{1}{2}}}{\rm d}t.
\end{displaymath} 
After integrating the above averaged equations of motion for
reasonable sets of parameters, using a 4th-order Runge-Kutta method with  variable
stepsize (Press et al. 1996), it was noticed that ${e_{{\rm T}}}$ remained
almost constant.  If that approximation is taken as an assumption, and
terms 
of order ${e^{2}_{{\rm S}}}$ are neglected and only the dominant term 
is retained in equation (\ref{gtd}) (the dominant term is proportional
to ${\beta}$, while the next order term is proportional to
${\alpha\beta}$, which, for the range of parameters discussed in this
paper, is rather small compared to the dominant term), then the system can be reduced to one that can be
solved analytically:
\begin{eqnarray}
\frac{{\rm d}x_{{\rm S}}}{{\rm d}\tau} & = & -By_{{\rm S}}+C\sin{g_{{\rm T}}}\nonumber\\
\frac{{\rm d}y_{{\rm S}}}{{\rm d}\tau} & = & Bx_{{\rm S}}-C\cos{g_{{\rm T}}}
\label{dior}\\
\frac{{\rm d}g_{{\rm T}}}{{\rm d}\tau} & = & A\nonumber,
\end{eqnarray} 
where
\begin{displaymath}
A=\frac{\beta}{(1-e^{2}_{{\rm T}})^{2}},\hspace{0.3cm}
B=\frac{1}{(1-e^{2}_{{\rm
T}})^{\frac{3}{2}}}+\frac{25}{8}\gamma\frac{3+2e^{2}_{{\rm
T}}}{(1-e^{2}_{{\rm T}})^{3}}\hspace{0.2cm} \mbox{and} \hspace{0.1cm}C=\frac{5}{4}\alpha\frac{e_{{\rm T}}}{(1-e^{2}_{{\rm
T}})^{\frac{5}{2}}}.
\end{displaymath}

The solution to system (\ref{dior}) is:
\begin{eqnarray}
x_{{\rm S}}(\tau) & = &
C_{1}\cos{B\tau}+C_{2}\sin{B\tau}+\frac{C}{B-A}\cos{(A\tau+g_{{\rm T}_{0}})}\label{secsol1} \\
y_{{\rm S}}(\tau) & = & C_{1}\sin{B\tau}-C_{2}\cos{B\tau}+\frac{C}{B-A}\sin{(A\tau+g_{{\rm T}_{0}})}\label{secsol2}
\end{eqnarray}
where
\begin{math}
C_{1}, C_{2}
\end{math}
are constants of integration and 
\begin{math}
{g_{{\rm T}}}_{0}
\end{math}
is the initial value of ${g_{{\rm T}}}$.

\subsection{A formula for the inner eccentricity}
\label{s2.3}

In sections (\ref{s1}) and (\ref{s2}) we derived expressions for the
short period and secular contribution to the inner eccentric vector.
These can be combined to give an expression for the total eccentricity by replacing the constants in equations (\ref{e11}) and
(\ref{e12}) by equations (\ref{secsol1}) and (\ref{secsol2}), since
the latter evolve on a much larger timescale.  This yields:
\begin{eqnarray}
x_{{\rm in}} & = & x_{1}(t)-C_{{\rm x}_{1}}+x_{{\rm S}}\label{xtot}\\
y_{{\rm in}} & = & y_{1}(t)-C_{{\rm y}_{1}}+y_{{\rm S}}\label{ytot}
\end{eqnarray}
The constants
${C_{1}}$ and ${C_{2}}$ in equations (\ref{secsol1}) and
(\ref{secsol2}) are determined by the fact that the inner eccentricity
is initially zero and are found to be
\begin{eqnarray}
C_{1} & = & -x_{1}(0)-\frac{C}{B-A}\\ 
C_{2} & = & y_{1}(0)+\frac{C}{B-A} 
\end{eqnarray}
We are now able to obtain an expression for the inner eccentricity.
Averaging over time and over the initial true anomaly ${f_{0}}$ and ${\varpi}$, the
averaged square inner eccentricity will be given by:

\begin{eqnarray}
\overline{e_{{\rm in}}^{2}} & = &
\frac{m_{3}^{2}}{M^{2}}\frac{1}{X^{4}(1-e^{2})^{\frac{9}{2}}}\left[\frac{43}{8}+\frac{129}{8}e^{2}+\frac{129}{64}e^{4}+\frac{1}{(1-e^{2})^{\frac{3}{2}}}(\frac{43}{8}+\frac{645}{16}e^{2}+\frac{1935}{64}e^{4}+\frac{215}{128}e^{6})+\frac{1}{X^{2}(1-e^{2})^{3}}[\frac{365}{18}+\right.\nonumber\\
& &
+\frac{44327}{144}e^{2}+\frac{119435}{192}e^{4}+\frac{256105}{1152}e^{6}+\frac{68335}{9216}e^{8}+\frac{1}{(1-e^{2})^{\frac{3}{2}}}(\frac{365}{18}+\frac{7683}{16}e^{2}+\frac{28231}{16}e^{4}+\frac{295715}{192}e^{6}+\frac{2415}{8}e^{8}+\nonumber\\
& &
+\frac{12901}{2048}e^{10})]+\frac{1}{X(1-e^{2})^{\frac{3}{2}}}[\frac{61}{3}+\frac{305}{2}e^{2}+\frac{915}{8}e^{4}+\frac{305}{48}e^{6}+\frac{1}{(1-e^{2})^{\frac{3}{2}}}(\frac{61}{3}+\frac{854}{3}e^{2}+\frac{2135}{4}e^{4}+\frac{2135}{12}e^{6}+\nonumber\\
& &
+\frac{2135}{384}e^{8})]+m_{*}^{2}X^{\frac{2}{3}}(1-e^{2})[\frac{225}{256}+\frac{3375}{1024}e^{2}+\frac{7625}{2048}e^{4}+\frac{29225}{8192}e^{6}+\frac{48425}{16384}e^{8}+\frac{825}{2048}e^{10}+\frac{1}{(1-e^{2})^{\frac{3}{2}}}(\frac{225}{256}+\nonumber\\
& &
+\frac{2925}{1024}e^{2}+\frac{775}{256}e^{4}+\frac{2225}{8192}e^{6}+\frac{25}{512}e^{8})]+m_{*}^{2}\frac{1}{X^{\frac{4}{3}}(1-e^{2})^{2}}[\frac{8361}{4096}+\frac{125415}{8192}e^{2}+\frac{376245}{32768}e^{4}+\frac{41805}{65536}e^{6}+\nonumber\\
& & \left.+\frac{1}{(1-e^{2})^{\frac{3}{2}}}(\frac{1575}{512}+\frac{11025}{256}e^{2}+\frac{165375}{2048}e^{4}+\frac{55125}{2048}e^{6}+\frac{55125}{65536}e^{8})]\right]+2(\frac{C}{B-A})^{2}.\label{fifo}
\end{eqnarray}

It should be mentioned here that in order to average the ${P_{31}}$
term over ${f}$ , we chose to expand ${(1+e\cos{f})^{-2}}$ binomially
including terms up to ${O(e^{3})}$.  It should also be mentioned that
numerical and theoretical tests showed that the total and secular
outer eccentric vectors were initially almost equal and therefore in
the evaluation of the above formula we consider ${e_{{\rm T}}=e}$ and 
${{g_{{\rm T}}}_{0}=\varpi}$.

\section{COMPARISON WITH NUMERICAL RESULTS}
In order to test the validity of the formulae derived in the previous
sections, we integrated the full equations of motion numerically, using
a symplectic integrator with time transformation (Mikkola 1997).

The code calculates the relative position and velocity vectors of the
two binaries at every time step.  Then, by using standard two body formulae,
we computed the orbital elements of the two binaries.
The various parameters used by the code, were given the following
values: writing index ${Iwr=1}$, average number of steps per inner binary
period ${NS=60}$, method coefficients ${a1=1}$ and ${a2=15}$,
correction index ${icor=1}$.
In all simulations, we confined ourselves to systems with mass ratios within the
range ${10:1}$ since, among stellar triples, 'mass ratios are rare
outside a range of approximately ${10:1}$, although such systems would be
inherently difficult to recognise' (Eggleton ${\&}$ Kiseleva 1995); 
We also introduced the fictitious initial period ratio ${X_{0f}}$, defined as
the ratio of the period that the outer binary would have on a circular
orbit with a semi major axis equal to its periastron distance over the
period of the inner binary.  In all cases ${X_{0f} \geq 10}$.  We also used units such that
${G=1}$ and ${m_{1}+m_{2}=1}$ and we always started the integrations
with ${a_{1}=1}$.  In that system of units, the initial conditions for the numerical
integrations were as follows:
\begin{displaymath}
r_{1}=1,\hspace{0.5cm} r_{2}=0,\hspace{0.5cm} r_{3}=0
\end{displaymath}	
\begin{displaymath}
R_{1}=R_{0}\cos{(f_{0}+\varpi)},\hspace{0.5cm} R_{2}=R_{0}\sin{(f_{0}+\varpi)},\hspace{0.5cm} R_{3}=0
\end{displaymath}	
\begin{displaymath}
\dot{r}_{1}=0,\hspace{0.5cm} \dot{r}_{2}=1,\hspace{0.5cm} \dot{r}_{3}=0
\end{displaymath}	
\begin{displaymath}
\dot{R}_{1}=-\sqrt{\frac{M}{a_{2}(1-e^{2})}}\sin{(f_{0}+\varpi)},\hspace{0.5cm} \dot{R}_{2}=\sqrt{\frac{M}{a_{2}(1-e^{2})}}\cos{(f_{0}+\varpi)},\hspace{0.5cm} \dot{R}_{3}=0.
\end{displaymath}

\subsection{SHORT PERIOD EVOLUTION} 

First we tested the validity of equations (\ref{xtot}) and
(\ref{ytot}) in the short term.  The results are presented in Tables 1
and 2, which give the percentage error between the
averaged, over time, numerical and theoretical ${e_{{\rm in}}}$ (the theoretical eccentricity was obtained by
evaluating  equations (\ref{xtot}) and (\ref{ytot}) everytime we
had an output from the symplectic integrator; both averaged
numerical and theoretical eccentricities  were calculated by using the
trapezium rule).  Table 1 presents results for ${e=0.4}$, while Table 2 presents results for
${e=0.75}$.  For each pair ${(m_{3},X_{0})}$ in these Tables, there are five entries,
corresponding, from top to bottom, to the following inner binaries:
${m_{1}=0.1-m_{2}=0.9}$, ${m_{1}=0.2-m_{2}=0.8}$,
${m_{1}=0.3-m_{2}=0.7}$, ${m_{1}=0.4-m_{2}=0.6}$ and
${m_{1}=0.5-m_{2}=0.5}$.  A dash denotes that the analogy among the masses was
outside the range ${10:1}$.  The
integrations were performed over one outer orbital period time span
and were done for ${f_{0}=90^{\circ}}$ and ${\varpi=0^{\circ}}$.  However, this does
not affect the qualitative understanding of the problem at all. 

The results generally show a good agreement between the numerical and
theoretical eccentricity. All the errors are below ${10\%}$ and they
drop as we move to larger values of ${X_{0f}}$.  Similar results are
obtained for different ${f_{0}}$ and ${\varpi}$.  However the theory
is not very accurate for highly eccentric outer binaries, because of
the expansion for ${f}$ given by equation (\ref{expa}).  In this case, higher
order terms with respect to the eccentricity should be included in
equation (\ref{expa}) for a better approach to the problem.  The masses of the
inner binary also play an important role in that specific issue, as
the ${P_{3}}$ term is proportional to (${m_{1}-m_{2}}$) and therefore
the closer we are in a situation where the inner binary has equal
masses the smaller the problem of neglecting terms of ${O(e^{4})}$ in
equation (\ref{expa}) will be.  Finally, the theory can work well even when
our short period model includes only the ${P_{21}}$ and ${P_{31}}$
terms.  The difference is more apparent for smaller ${X_{0f}}$ and
for smaller outer eccentricities where the full model works much better
in those cases.  It is also necessary when ${m_{1}=m_{2}}$. 
 
\begin{figure}
\begin{center}
\includegraphics[width=80mm,height=60mm]{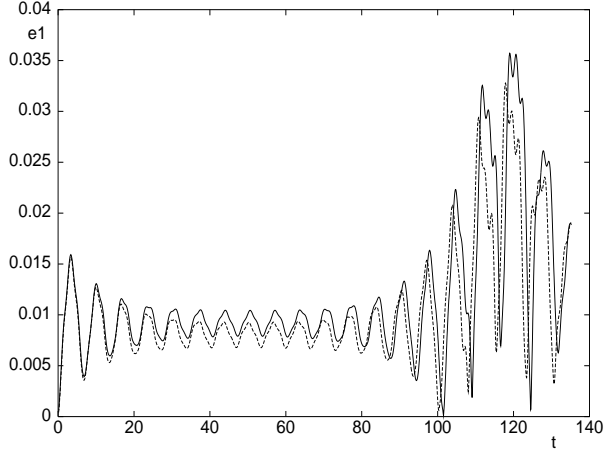}
\caption[]{Inner eccentricity against time for a system with
${m_{1}=0.3}$, ${m_{3}=3}$, ${e=0.4}$, ${X_{0f}=10}$, ${\varpi=0^{\circ}}$ and ${f_{0}=90^{\circ}}$.  The integration 
time span is one outer orbital period (${T_{{\rm out}}=135.2}$).  The
continuous curve comes from the numerical integration of the full equations of motion, while the
dashed curve is a plot of equations (\ref{xtot}) and (\ref{ytot}).  In
the system of units used, the inner binary period is ${T_{{\rm in}}=2\pi}$.}
\label{fig2}
\end{center}
\end{figure}

Fig. \ref{fig2} is a plot of inner binary eccentricity against time
for a system with ${m_{1}=0.3}$, ${m_{3}=3}$, ${e=0.4}$, ${X_{0f}=10}$, ${\varpi=0^{\circ}}$ and ${f_{0}=90^{\circ}}$.  The
continuous curve has been produced as a result of the numerical integration of
the full equations of motion, while the dashed curve is based on
equations (\ref{xtot}) and (\ref{ytot}).  The error for this case, as
seen in Table 1, was ${9.6\%}$.

\begin{table}
\caption[]{Percentage error between the
averaged numerical and averaged theoretical ${e_{{\rm in}}}$.  The
theoretical model is based on equations (\ref{xtot}) and
(\ref{ytot}).  For all systems, ${e=0.4}$, ${f_{0}=90^{\circ}}$ and ${\varpi=0^{\circ}}$.}
\vspace{0.1 cm}
\begin{center}	
{\small \begin{tabular}{c c c c c c c c c c}\hline
${m_{3}\backslash\ X_{0}}$ & ${10}$ & ${15}$ & ${20}$ & ${25}$ & ${30}$ & ${50}$ \\
\hline
0.05 & - & - & - & - & - & - \\
     & - & - & - & - & - & - \\
     & - & - & - & - & - & - \\
     & - & - & - & - & - & - \\
     & 2.6 & 1.2 & 0.8 & 0.5 & 0.4 & 0.2 \\
0.09 & 3.5 & 1.9 & 1.3 & 1 & 0.8 & 0.5 \\
     & 3.3 & 1.8 & 1.2 & 0.9 & 0.7 & 0.4 \\
     & 3.2 & 1.6 & 1.1 & 0.8 & 0.6 & 0.3 \\
     & 3 & 1.5 & 1 & 0.7 & 0.5 & 0.3 \\
     & 2.8  & 1.4 & 0.9 & 0.6 & 0.5 & 0.3 \\
0.5  & 6.5 & 4.1 & 3.1 & 2.4 & 2 & 1.3 \\
     & 6 & 3.6 & 2.7 & 2 & 1.7 & 1.1 \\
     & 5.5 & 3.2 & 2.3 & 1.7 & 1.4 & 0.9 \\
     & 5.1 & 3 & 2.1 & 1.6 & 1.2 & 0.7 \\
     & 4.9 & 2.8 & 2 & 1.4 & 1.1 & 0.6 \\
 1   & 8.4 & 5.3   & 4 & 3.3 & 2.7 & 1.7 \\
     & 7.7  & 4.8  & 3.5 & 2.8 & 2.3 & 1.5  \\
     & 7.1   & 4.3  & 3.1 & 2.4 & 2 & 1.2  \\
     & 6.6 & 4 & 2.8 & 2.1 & 1.7 & 1  \\
     & 6.4 & 3.8  & 2.7  & 1.9  & 1.5 & 0.9 \\
1.5  & - & - & - & - & - & - \\
     & 8.7 & 5.4 & 4.1 & 3.1 & 2.6 & 1.6 \\
     & 8.1 & 5 & 3.7 & 2.7 & 2.3 & 1.3 \\
     & 7.6 & 4.6 & 3.3 & 2.4  & 2 & 1.1 \\
     & 7.3 & 4.4 & 3.1 & 2.2  & 1.8 & 1 \\
 2   & - & - & - & - & - & - \\
     & 9.3 & 5.8 & 4.3 & 3.4 & 2.8 & 1.7 \\
     & 8.7 & 5.3 & 3.9 & 3 & 2.4 & 1.4 \\
     & 8.2 & 5  & 3.5 & 2.7 & 2.1 & 1.2 \\
     & 7.8 & 4.8 & 3.4 & 2.5 & 2 & 1.1 \\
2.6  & - & - & - & - & - & - \\
     & - & - & - & - & - & - \\
     & 9.3 & 5.6 & 4.2 & 3.1  & 2.5 & 1.5 \\
     & 8.8 & 5.3 & 3.8 & 2.8 & 2.3 & 1.3 \\
     & 8.4 & 5.2 & 3.7 & 2.6 & 2.1 & 1.2 \\
 3   & - & - & - & - & - & - \\
     & - & - & - & - & - & - \\
     & 9.6 & 5.8 & 4.2 & 3.2 & 2.5  & 1.5 \\
     & 9 & 5.5 & 3.9 & 2.9 & 2.4 & 1.3 \\
     & 8.7 & 5.3 & 3.8 & 2.8  & 2.2 & 1.2 \\
3.4  & - & - & - & - & - & - \\
     & - & - & - & - & - & - \\
     & - & - & - & - & - & - \\
     & 9.3 & 5.6 & 4 & 3 & 2.4 & 1.4 \\
     & 8.9 & 5.5 & 3.9  & 2.8  & 2.2  & 1.3 \\
 4   & - & - & -  & - & - & - \\
     & - & - & - & - & - & - \\
     & - & - & - & - & - & - \\
     & 9.6 & 5.7 & 4.2 & 3.1 & 2.5 & 1.4 \\
     & 9.3 & 5.6 & 4 & 2.9 & 2.4 & 1.3 \\
4.5  & - & - & - & - & - & - \\
     & - & - & - & - & - & - \\
     & - & - & - & - & - & - \\
     & - & - & - & - & - & - \\
     & 9.5 & 5.7 & 4.1 & 3 & 2.3 & 1.3 \\
 5   & - & - & - & - & - & - \\
     & - & - & - & - & - & - \\
     & - & - & - & - & - & - \\
     & - & - & - & - & - & - \\
     & 9.7 & 5.8 & 4.2 & 3 & 2.4 & 1.4 \\

\hline
\end{tabular}}
\end{center}	
\end{table}

\begin{table}
\caption[]{Percentage error between the
averaged numerical and averaged theoretical ${e_{{\rm in}}}$.  The
theoretical model is based on equations (\ref{xtot}) and
(\ref{ytot}).  For all systems, ${e=0.75}$, ${f_{0}=90^{\circ}}$ and ${\varpi=0^{\circ}}$.}
\vspace{0.1 cm}
\begin{center}	
{\small \begin{tabular}{c c c c c c c c c c}\hline
${m_{3}\backslash\ X_{0}}$ & ${10}$ & ${15}$ & ${20}$ & ${25}$ & ${30}$ & ${50}$ \\
\hline
0.05 & - & - & - & - & - & - \\
     & - & - & - & - & - & - \\
     & - & - & - & - & - & - \\
     & - & - & - & - & - & - \\
     & 0.8 & 0.4 & 0.3 & 0.2 & 0.1 & 0.1 \\
0.09 & 1.1 & 0.4 & 0.1 & -0.2 & -0.4 & -1 \\
     & 1.7 & 0.6 & 0.3 & 0.1 & -0.1 & -0.4 \\
     & 1.8 & 0.7 & 0.4 & 0.3 & 0.2 & -0.1 \\
     & 1.5 & 0.6 & 0.4 & 0.3 & 0.2 & 0.1 \\
     & 1 & 0.5 & 0.4 & 0.3 & 0.2 & 0.1 \\
0.5  & 2.3 & 1.2 & 0.7 & 0.4 & 0.1 & -0.5 \\
     & 2.9 & 1.4 & 0.9 & 0.6 & 0.4 & -0.1 \\
     & 3 & 1.5 & 1 & 0.7 & 0.6 & 0.2 \\
     & 2.9 & 1.5 & 1.1 & 0.8 & 0.6 & 0.3 \\
     & 2.7 & 1.5 & 1.1 & 0.9 & 0.7 & 0.4 \\
 1   & 3.1 & 1.7  & 1.1 & 0.7 & 0.5 & -0.1 \\
     & 3.6  & 1.8 & 1.2 & 0.9 & 0.7 & 0.2  \\
     & 3.7 & 2 & 1.4 & 1 & 0.8 & 0.4  \\
     & 3.8 & 2.1 & 1.5 & 1.1 & 0.9 & 0.5  \\
     & 3.7 & 2.2  & 1.6  & 1.3 & 1 & 0.6 \\
1.5  & - & - & - & - & - & - \\
     & 4 & 2.1 & 1.5 & 1.1 & 0.8 & 0.3 \\
     & 4.2 & 2.3 & 1.6 & 1.2 & 1 & 0.5 \\
     & 4.3 & 2.4 & 1.7 & 1.3  & 1.1 & 0.6 \\
     & 4.3 & 2.6 & 1.9 & 1.5  & 1.2 & 0.7 \\
 2   & - & - & - & - & - & - \\
     & 4.2 & 2.3 & 1.6 & 1.2 & 0.9 & 0.4 \\
     & 4.5 & 2.5 & 1.8 & 1.4 & 1.1 & 0.6 \\
     & 4.7 & 2.6 & 1.9 & 1.5 & 1.2 & 0.7 \\
     & 4.8 & 2.9 & 2.1 & 1.7 & 1.4 & 0.8 \\
2.6  & - & - & - & - & - & - \\
     & - & - & - & - & - & - \\
     & 4.8 & 2.7 & 1.9 & 1.5  & 1.2 & 0.6 \\
     & 4.9 & 2.9 & 2.1 & 1.6 & 1.3 & 0.7 \\
     & 5.1 & 3.1 & 2.3 & 1.8 & 1.5 & 0.9 \\
 3   & - & - & - & - & - & - \\
     & - & - & - & - & - & - \\
     & 4.9 & 2.8 & 2 & 1.5 & 1.2  & 0.7 \\
     & 5.1 & 3 & 2.1 & 1.7 & 1.4 & 0.8 \\
     & 5.3 & 3.2 & 2.4 & 1.9 & 1.5 & 0.9 \\
3.4  & - & - & - & - & - & - \\
     & - & - & - & - & - & - \\
     & - & - & - & - & - & - \\
     & 5.3 & 3 & 2.2 & 1.7 & 1.4 & 0.8 \\
     & 5.4 & 3.3 & 2.4  & 1.9  & 1.6  & 0.9 \\
 4   & - & - & -  & - & - & - \\
     & - & - & - & - & - & - \\
     & - & - & - & - & - & - \\
     & 5.4 & 3.2 & 2.3 & 1.8 & 1.5 & 0.8 \\
     & 5.6 & 3.4 & 2.5 & 2 & 1.6 & 1 \\
4.5  & - & - & - & - & - & - \\
     & - & - & - & - & - & - \\
     & - & - & - & - & - & - \\
     & - & - & - & - & - & - \\
     & 5.7 & 3.5 & 2.6 & 2 & 1.7 & 1 \\
 5   & - & - & - & - & - & - \\
     & - & - & - & - & - & - \\
     & - & - & - & - & - & - \\
     & - & - & - & - & - & - \\
     & 5.9 & 3.6 & 2.6 & 2 & 1.7 & 1 \\

\hline
\end{tabular}}
\end{center}	
\end{table}

\subsection{LONG PERIOD EVOLUTION}

Next, we tested equation (\ref{fifo}), which accounts for the short
period and secular effects to the inner eccentricity.  The formula was
compared with results obtained from integrating the full
equations of motion numerically.  These results are presented in Table
3, which gives the percentage error between the averaged (over time,
initial true anomaly ${f_{0}}$ and ${\varpi}$) numerical ${e^{2}_{{\rm
in}}}$ and equation (\ref{fifo}).  The error is accompanied by the
period of the oscillation of the eccentricity, which is the same as
the integration time span.  There are four values per
(${m_{3}-X_{0f}}$) pair.  The first two (error-period) correspond to a
system with ${e=0.4}$, while the other two to a system with
${e=0.75}$.
  
Each system was numerically integrated for
${\varpi=0^{\circ}-360^{\circ}}$ and ${f_{0}=0^{\circ}-360^{\circ}}$
with a step of ${45^{\circ}}$.  For a given value of ${\varpi}$ and
${f_{0}}$ we integrated our system. After
each run, ${e^{2}_{{\rm in}}}$ was averaged over time using the trapezium
rule and then we integrated the system for a different ${f_{0}}$.  After the integrations for all ${f_{0}}$ were done, we
averaged over ${f_{0}}$ by using the rectangle rule. Then, the same
procedure was applied for the next value of ${\varpi}$ and when the integrations
for all ${\varpi}$ were done, we averaged over ${\varpi}$ by using the
rectangle rule.   The integrations
were also done for smaller steps in ${\varpi}$ and ${f_{0}}$, but there was not any difference in the outcome.
All the integrations presented in Table 3 were done for ${m_{1}=0.2}$,
but similar results are expected for the other inner
binary masses (note that for ${m_{1}=m_{2}}$ there is not any long
period oscillation in the inner eccentricity, as seen in section \ref{s2}). 

Generally, it appears that the theory is in agreement with the
numerical integrations.  There are some cases where the eccentricity of
the inner binary reaches significant values over a long period (e.g. for ${X_{0f}=10}$,
${m_{3}=0.09}$ and ${e=0.75}$ the inner maximum eccentricity is about
0.35) and therefore terms of ${O(e^{2}_{{\rm S}})}$ should be
included in the secular equations.  The contribution from short period terms becomes more
noticeable and hence increasingly important as the outer eccentricity
drops; and also as ${m_{3}}$ increases.  Numerical integrations were also performed for ${10}$ and ${100}$ secular periods without any change in the error (something expected when the error in the rate of change of the eccentricity is quasi-periodic).  Finally, we compared the results for the systems of
Table 3 with a formula which only included the ${P_{21}}$ and
${P_{31}}$ terms.  The maximum difference in the errors was around ${3\%}$.

\begin{table}
\caption[]{Percentage error between the averaged numerical
${e^{2}_{{\rm in}}}$ and equation (\ref{fifo}).}
\vspace{0.1 cm}
\begin{center}	
{\footnotesize \begin{tabular}{c c c c}\hline
${m_{3}\backslash\ X_{0f}}$ & ${10}$ & ${20}$ & ${30}$ \\
\hline
0.09 &-11.6&-2.5&-1 \\
     &106000&313000&625000 \\
     &22&2.3&1 \\
     &595000&1485000&2980000 \\
0.5  &-3.7&3.7&3.5 \\
     &7700 &35500&82500 \\
     &-3.9&1&2 \\
     &40000&185000&427000 \\
 1   &1&7.8&5.1 \\
     &4300&21000&51000 \\
     &-8.3&2.2&3 \\
     &23500&112000&265000 \\
1.5  &2&7.8&6.5 \\
     &3400 &17000 &41000 \\
     &-8.6&3.3&3.6 \\
     &18000&89000&214000 \\
 2   &10.5&11.2&7.7 \\
     &2700 &14500 &36000 \\
     &-5.3&3.9&3.5 \\
     &15000&78000&190000 \\
\hline
\end{tabular}}
\end{center}	
\end{table}

\section{Discussion}

We have derived a formula which gives an estimate for the inner eccentricity in
hierarchical triple systems with eccentric outer binaries and coplanar
orbits.  The theoretical model appears to work satisfactory for the
parameter ranges discussed.  In cases with large outer eccentricities
and significantly different inner binary masses, due to the
approximation of ${(1+e\cos{f})^{-2}}$ with a series expansion in terms
of ${e}$, the model could be inaccurate, especially for describing
short term evolution.  However, for hierarchical triple systems with
highly eccentric outer binaries, it appears that the inner eccentricity is dominated by secular evolution, which is not
affected by that approximation (the only effect of that approximation
will be on determining the constants of integration ${C_{1}}$ and
${C_{2}}$ in section \ref{s2.3}).  A shorter formula can also be derived by only retaining the ${P_{21}}$
and ${P_{31}}$ terms in equations (\ref{xtot}) and (\ref{ytot}).  As the numerical
integrations demonstrated in the previous section, the omission of
those terms did not affect the situation very much.  However, the
contribution of  the ${P_{22}}$ and ${P_{32}}$ terms is important in
cases with significant short period evolution, i.e. small outer
eccentricity and strong perturbation to the inner binary.

The theoretical results obtained in the previous sections could be applied in various astronomical and astrophysical topics.  As it was stated in the introduction, the separation of the components in {\sl close} binary systems can play a vital role in their evolution.  For instance, the flow of material in a semidetached system can be seriously affected by a small change in the separation of the two stars and therefore it would be useful to have some information about the eccentricity injected into the binary by the third star.  The theory could also be used in observational astronomy, to put constrains on differrent parameters of the observed systems, since there is always some uncertainty in their determination.  For example, the theory could be used to rule whether the orbits of the triple system are coplanar or not, an interesting question which is related to the formation of the system.  

The same technique can be applied to investigate systems with non coplanar orbits.

\section*{ACKNOWLEDGMENTS}
The author is grateful to Prof. Douglas Heggie for all the useful
discussions on the context of this paper. I also want to thank Seppo
Mikkola, who kindly provided the code for integrating hierarchical
triple systems.
 
\section*{REFERENCES}
 
Eggleton P. P., Kiseleva L. G., 1995, ApJ, 455, 640\\
Fekel F. C., Jr.; Tomkin J., 1982, ApJ 263, 289\\
Georgakarakos N., 2002, MNRAS, 337, 559\\
Hinkle K. H., Fekel F. C., Johnson D. S., Scharlach W. W. G., 1993, AJ, 105, 1074\\
Jha S., Torres G., Stefanik R. P., Latham D. W., Mazeh T., 2000, MNRAS, 317, 375\\
Kiseleva L. G., Eggleton P. P., Mikkola S., 1998, MNRAS, 300, 292\\
Marchal C., 1990, The Three-Body Problem.  Elsevier Science Publishers, the Netherlands\\
Mikkola S., 1997, CeMDA, 67, 145\\
Murray C. D., Dermott S. F., 1999, Solar System Dynamics. Cambridge
Univ. Press, Cambridge\\
Press W. H., Teukolsky S. A., Vetterling W. T.,
Flannery B. P., 1996, Numerical Recipes In Fortran 77 (2nd
ed.).  Cambridge Univ. Press, NY
\end{document}